\documentclass[a4paper]{jpconf}

\usepackage{graphicx}
\usepackage{amsmath,amssymb,amsfonts,amscd}
\usepackage{xspace}

\newcommand{\modl}[1]{model \texttt{#1}}
\newcommand{\modls}[1]{models \texttt{#1}}
\newcommand{\Modl}[1]{Model \texttt{#1}}

\newcommand{\modelname}[1]{\texttt{#1}}

\newcommand{\eg}{e.g.\xspace}

\newcommand{\msol}{M_{\odot}}

\newcommand{\km}{\textrm{km}}

\newcommand{\sek}{\textrm{s}}

\newcommand{\Gauss}{\textrm{G}}

\newcommand{\figref}[1]{Fig.\,\ref{#1}}

\newcommand{\secref}[1]{Sect.\,\ref{#1}}

\newcommand{\nat}{Nature}

\newcommand{\apj}{ApJ}
\newcommand{\mnras}{MNRAS}

\newcommand{\aap}{A{\&}A}

\newcommand{\apjl}{ApJL}

\newcommand{\pasj}{Publications of the ASJ}

\begin{document}

\title{Evolution of the surface magnetic field of rotating proto-neutron stars}

\author{M Obergaulinger and M{\'A} Aloy}

\address{%%
  Departamento de Astronom\'{\i}a y Astrof\'{\i}sica,  Universidad de Valencia,  C/ Dr.~Moliner 50, 46100 Burjassot, Spain
}

\ead{martin.obergaulinger@uv.es}

\begin{abstract}
  We study the evolution of the field on the surface of proto-neutron
  stars in the immediate aftermath of stellar core collapse by
  analyzing the results of self-consistent, axisymmetric simulations
  of the cores of rapidly rotating high-mass stars.  To this end, we
  compare the field topology and the angular spectra of the poloidal
  and toroidal field components over a time of about one seconds for
  cores. Both components are characterized by a complex geometry with
  high power at intermediate angular scales.  The structure is mostly
  the result of the accretion of magnetic flux embedded in
  the matter falling through the turbulent post-shock layer onto the
  PNS. Our results may help to guide further studies of the long-term
  magneto-thermal evolution of proto-neutron stars. We find that the
  accretion of stellar progenitor layers endowed with low or null
  magnetization bury the magnetic field on the PNS surface very
  effectively.
\end{abstract}

\section{Introduction}
\label{Sek:Intro}

The wide range of surface magnetic field strengths found across the
population of neutron stars
(\eg \cite{Kaspi__2010__ProceedingsoftheNationalAcademyofScience__Grandunificationofneutronstars})
is the result of a combination of processes operating at their
formation during stellar core collapse and effects that modify their
structure afterwards, such as cooling or accretion. Most observations
pertain to relatively old neutron stars, and do not place tight
constraints on the very early evolution of the field.  Hence, it is
difficult to draw conclusions on the fields of nascent proto-neutron
stars (PNSs) from observations or, conversely, to connect theoretical
results from models of supernova core collapse to evolved neutron
stars.  Nevertheless, the field is likely to bear at least a
significant imprint of these early conditions.

A thorough theoretical study of the evolution of the magnetic field of
PNSs should optimally be based on three-dimensional long-term
simulations including magnetohydrodynamics and neutrino transport,
which treat the global dynamics of core collapse and a possible
supernova explosion and the generation of the PNS field in a
self-consistent manner.  The enormous computational costs of spectral
neutrino transport mean that thus far only a small number of such
simulations exist with the most sophisticated ones
\cite{Winteler_et_al__2012__apjl__MagnetorotationallyDrivenSupernovaeastheOriginofEarlyGalaxyr-processElements,Mosta_et_al__2014__apjl__MagnetorotationalCore-collapseSupernovaeinThreeDimensions,Moesta_et_al__2015__nat__Alarge-scaledynamoandmagnetoturbulenceinrapidlyrotatingcore-collapsesupernovae}
run for only a fairly limited period.

Hence, less expensive axisymmetric models are still of considerable
use for approaching this topic
(\eg \cite{Burrows_etal__2007__ApJ__MHD-SN,Bisnovatyi-Kogan_Moiseenko__2008__ProgressofTheoreticalPhysicsSupplement__Core-CollapseSupernovae:MagnetorotationalExplosionsandJetFormation,Takiwaki_Kotake_Sato__2009__apj__Special_Relativistic_Simulations_of_Magnetically_Dominated_Jets_in_Collapsing_Massive_Stars,Harikae_et_al__2009__apj__Long-Term_Evolution_of_Slowly_Rotating_Collapsar_in_SRMHD,Sawai_et_al__2013__apjl__GlobalSimulationsofMagnetorotationalInstabilityintheCollapsedCoreofaMassiveStar}).
Depending on the rotational energy and the seed field of the
pre-collapse star, but also potentially on the input physics and the
numerical method and grid resolution, their results range from minor
modifications of non-magnetized versions of the same cores to
magnetically driven explosions of a preferentially axial
morphology. Depending on the global evolution, the PNS may accrete
matter at all latitudes or only through narrow streams with a
potential final collapse to a black hole (BH).  The time evolution of
the rate and geometry of mass accretion and the magnetization of the
accreted layers have an important impact on the accumulation of
magnetic flux at the PNS surface.  The picture thus generated may show
a shell of enhanced magnetic field strength, but it is also possible
for the magnetic field to be buried underneath additional layers of
weakly magnetized gas \cite{Torres-Forne_et_al__2016__mnras__Arepulsarsbornwithahiddenmagneticfield}.  These processes are complemented by the
possible amplification of magnetic fields in the interior of the PNS,
in particular if rotation and convection constitute a dynamo.

Previously, we studied the evolution of magnetized core collapse for
stars with an initial mass of $15 \, \msol$ without rotation
\cite{Obergaulinger_et_al__2014__mnras__Magneticfieldamplificationandmagneticallysupportedexplosionsofcollapsingnon-rotatingstellarcores}.
Varying the pre-collapse field strength between negligible values
below $10^{10} \, \Gauss$ and dynamically relevant, yet from the point
of view of stellar evolution rather overestimated, values of $10^{12}
\, \Gauss$, we found explosions driven by neutrino heating with
potentially a strong contribution of magnetic forces.  We were
following the evolution of the PNSs for several hundreds of
milliseconds and found the development of layers of strong field on
their surfaces owing their origin to the accretion of magnetized gas.
We now focus on cores of higher-mass stars with rapid rotation and
study the geometry of the magnetic field during the first up to two
seconds after PNS formation
\cite{Obergaulinger_Aloy__2017__mnras__Protomagnetarandblackholeformationinhigh-massstars}.
We note that our set of models includes both PNSs that collapse to a
BH within this time and ones that will likely remain stable for much
longer or potentially indefinitely.

We outline our simulations in \secref{Sek:Models}, discuss the result
in \secref{Sek:Res}, and present our conclusions in \secref{Sek:Con}.

\section{Simulations}
\label{Sek:Models}

Our numerical models of the stellar cores are based on special
relativistic magnetohydrodynamics (MHD), an approximately general
relativistic gravitational potential, and spectral neutrino transport.
Neutrino physics is treated in the hyperbolic two-moment formulation,
which provides a good approximation to a full solution of the
Boltzmann equation of radiative transfer and is intrinsically
multi-dimensional, and includes corrections due to the gas velocity
(\eg, Doppler shifts) and gravity and all relevant interactions
between neutrinos of all flavours and nucleons, nuclei, and electrons
and positrons
\cite{Just_et_al__2015__mnras__Anewmultidimensionalenergy-dependenttwo-momenttransportcodeforneutrino-hydrodynamics}.
The transport solver, computationally the most expensive part of the
simulations, is the main factor restricting us to axisymmetric rather
than full three-dimensional simulations.

As reported in
\cite{Obergaulinger_Aloy__2017__mnras__Protomagnetarandblackholeformationinhigh-massstars},
we simulated the core collapse of several stars with a high initial
mass of $35 \, \msol$ and subsolar metallicity (pre-collapse models
\modelname{35OB} and \modelname{35OC} of
\cite{Woosley_Heger__2006__apj__TheProgenitorStarsofGamma-RayBursts}).
To test whether these properties make them viable progenitors of long
GRBs, we combined the progenitors with different rotational profiles
and magnetic fields.  The results can be briefly summarized as
follows:
\begin{itemize}
\item In spite of their high compactness, usually taken as
  an indication for a high resistance to shock revival
  \cite{OConnor_Ott__2011__apj__BlackHoleFormationinFailingCore-CollapseSupernovae},
  our cores develop successful supernova explosions predominantly
  driven either by neutrino heating or by magnetorotational stresses.
  All explosions have a bipolar geometry with the magnetically driven
  showing the highest degree of collimation of the ejecta and the
  highest explosion energies.
\item While the ejecta expand asymmetrically, matter continues to fall
  onto the PNS, which therefore grows in mass.  For most of our
  models, the PNS reaches the threshold for BH formation within less than 2
  s after bounce.  Some models within this class show conditions that
  make a later collapsar phase conceivable.  
\item Models without a BH collapse tend to contain a rapidly rotating
  and hence notably flattened PNS of relatively strong magnetization.
\end{itemize}

\begin{figure}
  \centering
  \includegraphics[width=0.9\linewidth]{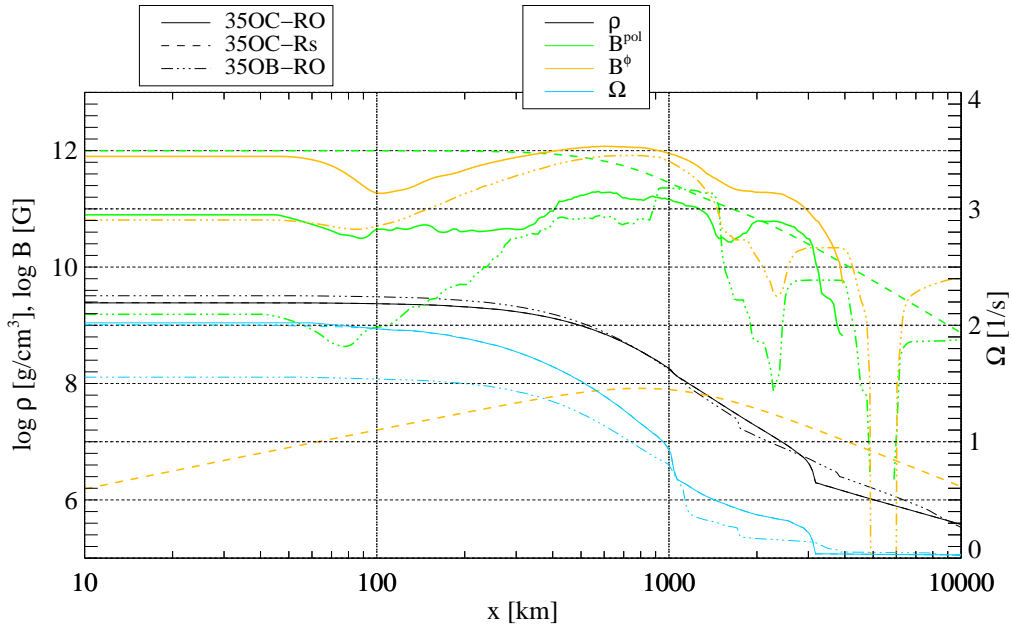}
  \caption{%%
    Profiles of density, angularly averaged poloidal and toroidal magnetic field, and
    angular velocity for all three models at the onset of collapse.
  }
  \label{Fig:initialmodels}
\end{figure}

In the following, we will present a study of the magnetic fields on
the surfaces of the PNSs of a selection of cores.  We focus on both
models that collapse to BHs and models where the PNS remains stable
throughout the simulation.  Although the simulations end too early for
observing the crust formation
\cite{Suwa__2014__pasj__Fromsupernovaetoneutronstars} and the late
time fall back of magnetized gas
\cite{Torres-Forne_et_al__2016__mnras__Arepulsarsbornwithahiddenmagneticfield},
we intend to get insights on the field structure of the very young
neutron stars.  To this end, we focus on the time evolution of the
strengths of the poloidal and toroidal components of the surface field
and analyze their multipole expansion in spherical harmonics in order
to distinguish large-scale structures from the field fluctuating on
small scales.  We refer to \figref{Fig:initialmodels} for the
pre-collapse state of the models, which, as the blue lines indicate,
all rotate rapidly and differentially.  We briefly describe their
dynamics in the following:
\begin{description}
\item[\Modl{35OC-RO}] includes the original rotational profile and
  magnetic field distribution of the core as given by the stellar
  evolution models.  The field has roughly similarly strong poloidal
  and toroidal components and is confined to the radiative zones of
  the core.  About 200 ms after core bounce, the model develops an
  explosion driven by a combination of magnetic forces and neutrino
  heating.  The ejecta expand in the form of bipolar outflows along
  the rotational axis while matter falls onto the PNS, whose mass and
  rotational energy continue to grow.  The accretion, however, does
  not reach the threshold of collapse to a BH.
\item[\Modl{35OC-Rs}] has a strong initial magnetic field based on the
  vector potential of
  \cite{Suwa_etal__2007__pasj__Magnetorotational_Collapse_of_PopIII_Stars}
  with the normalization set such that the profiles of toroidal and
  poloidal are equal and launches a magnetically powered, very
  energetic explosion almost immediately after bounce.  The explosion
  quenches the accretion onto the PNS rather effectively, leading to a
  reduced PNS mass and possibly preventing BH formation indefinitely.
  Internal redistribution of angular momentum from the centre to the
  outer layers makes the PNS assume an extremely oblate shape.
\item[\Modl{35OB-RO}] has the much weaker and strongly toroidally
  dominated magnetic field of the progenitor star \modelname{35OB},
  mostly eliminating the magnetic contribution to shock revival.  The
  higher compactness of the pre-collapse core causes the PNS to exceed
  the instability threshold and collapse to a BH at $t \sim 1.5 \,
  \sek$.
\end{description}

The magnetic field distribution of the stellar evolution models
\modelname{35OB/C} is the result of a simple model for the
amplification of the field by a dynamo operating only in radiative zones of
the stars \cite{Spruit__2002__AA__Dynamo}.  The  cores do
not include any magnetic field in convective zones.  Consequently,
there are large field-free regions in the cores, of which the ones
outside the iron core of \modl{35OB} are of particular importance, as
we will show below.

\section{Results}
\label{Sek:Res}

\begin{figure}
  \centering
  \includegraphics[width=0.32\linewidth]{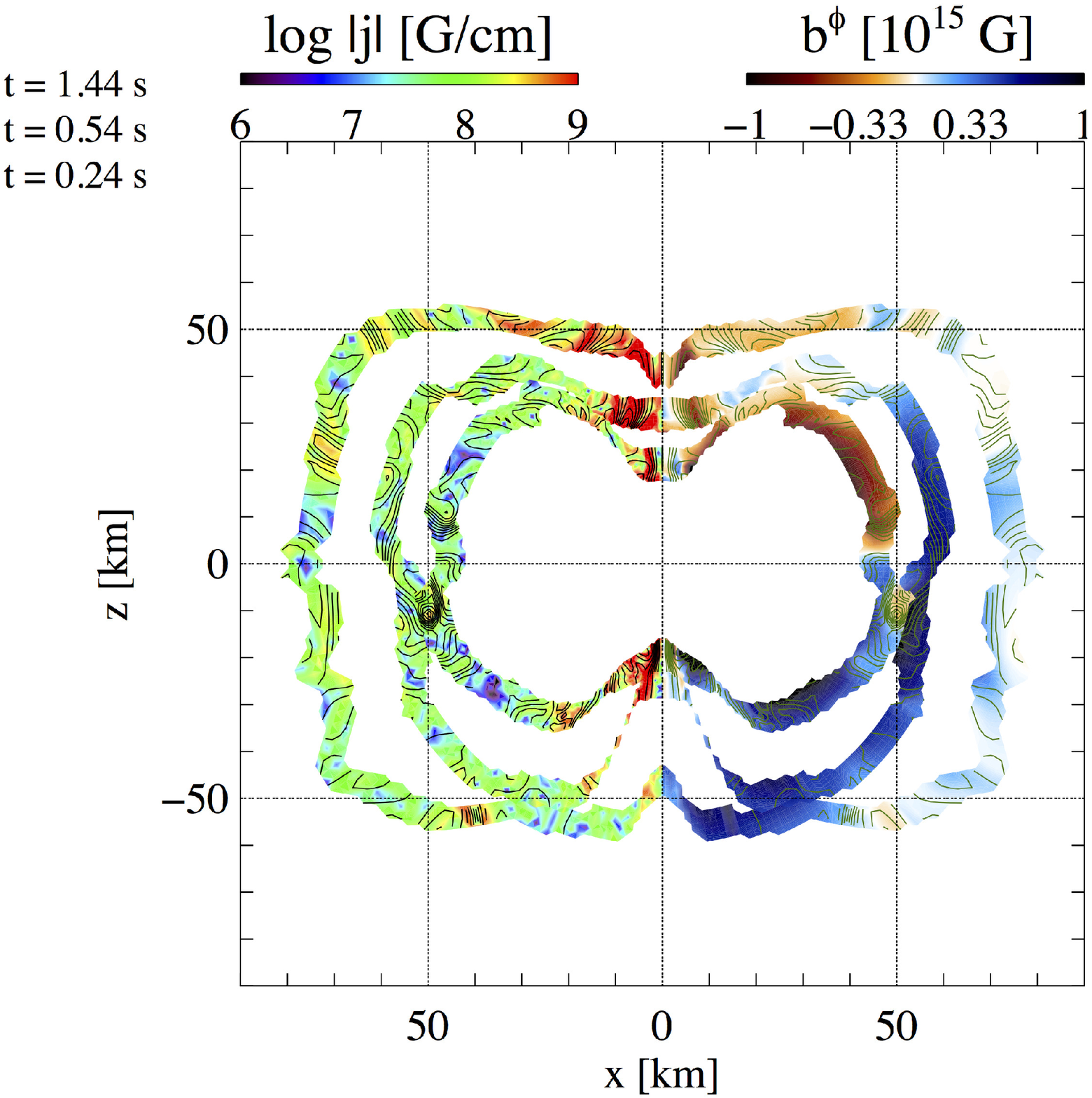}
  \includegraphics[width=0.32\linewidth]{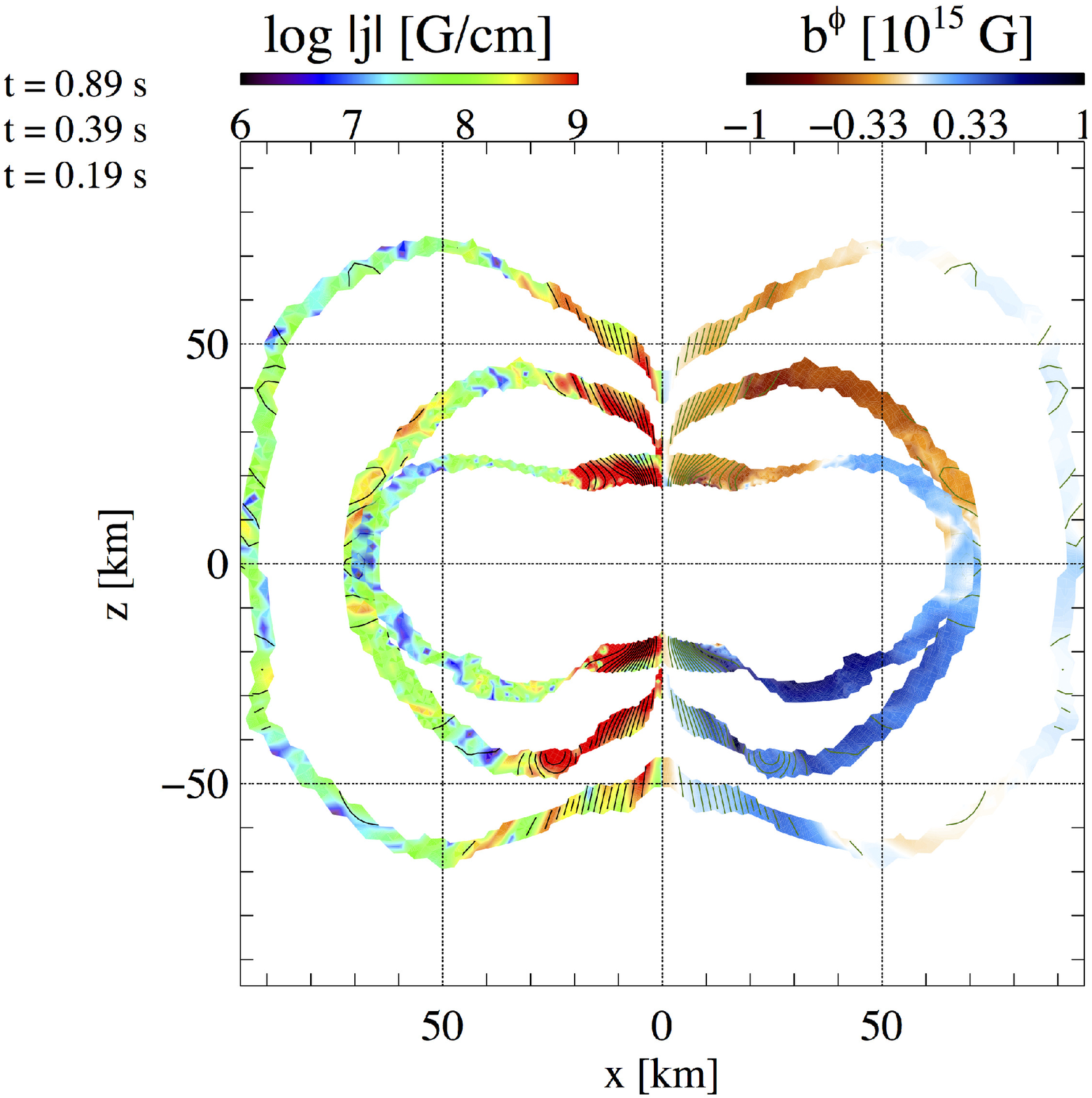}
  \includegraphics[width=0.32\linewidth]{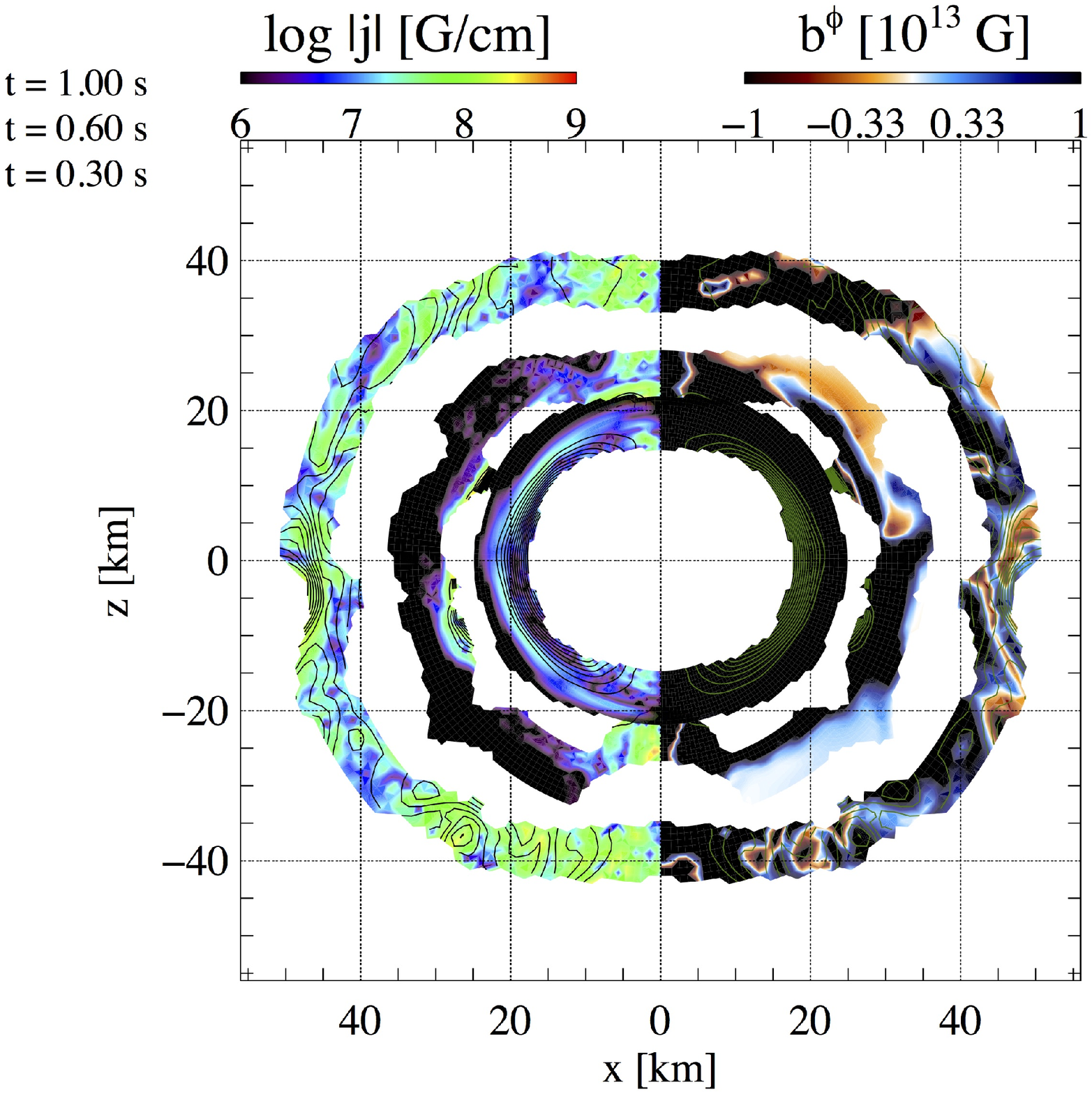}
  \includegraphics[width=0.32\linewidth]{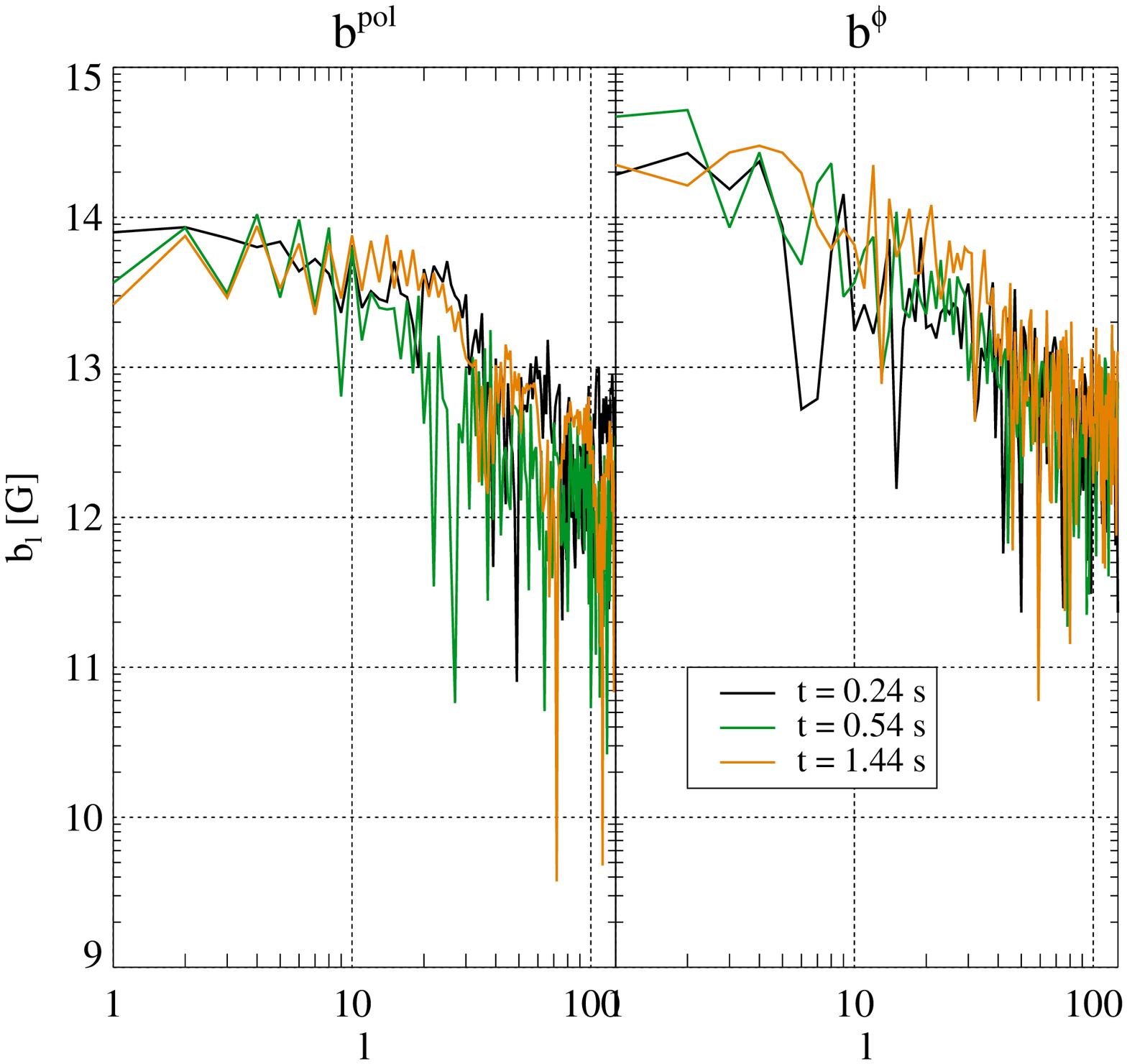}
  \includegraphics[width=0.32\linewidth]{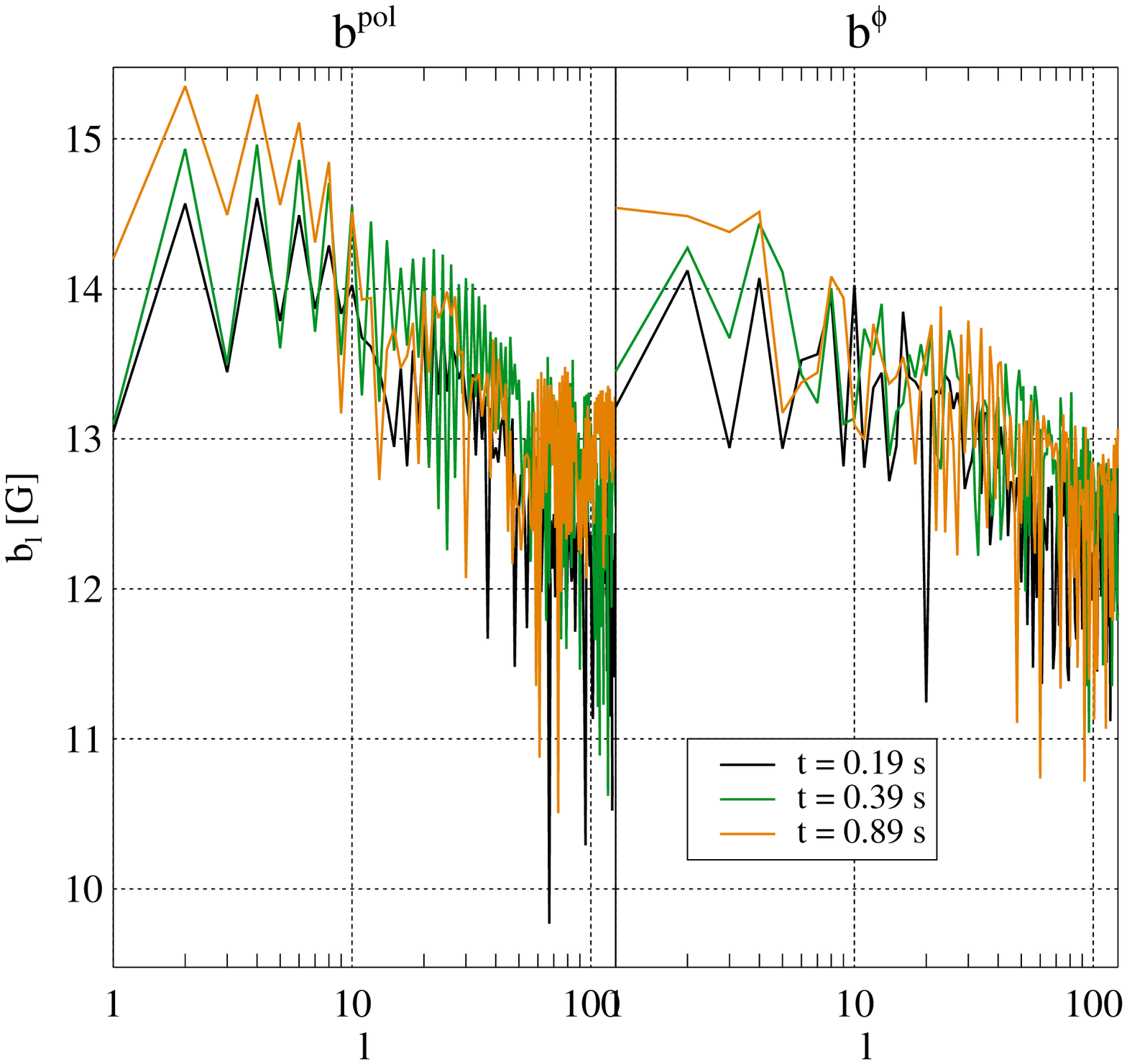}
  \includegraphics[width=0.32\linewidth]{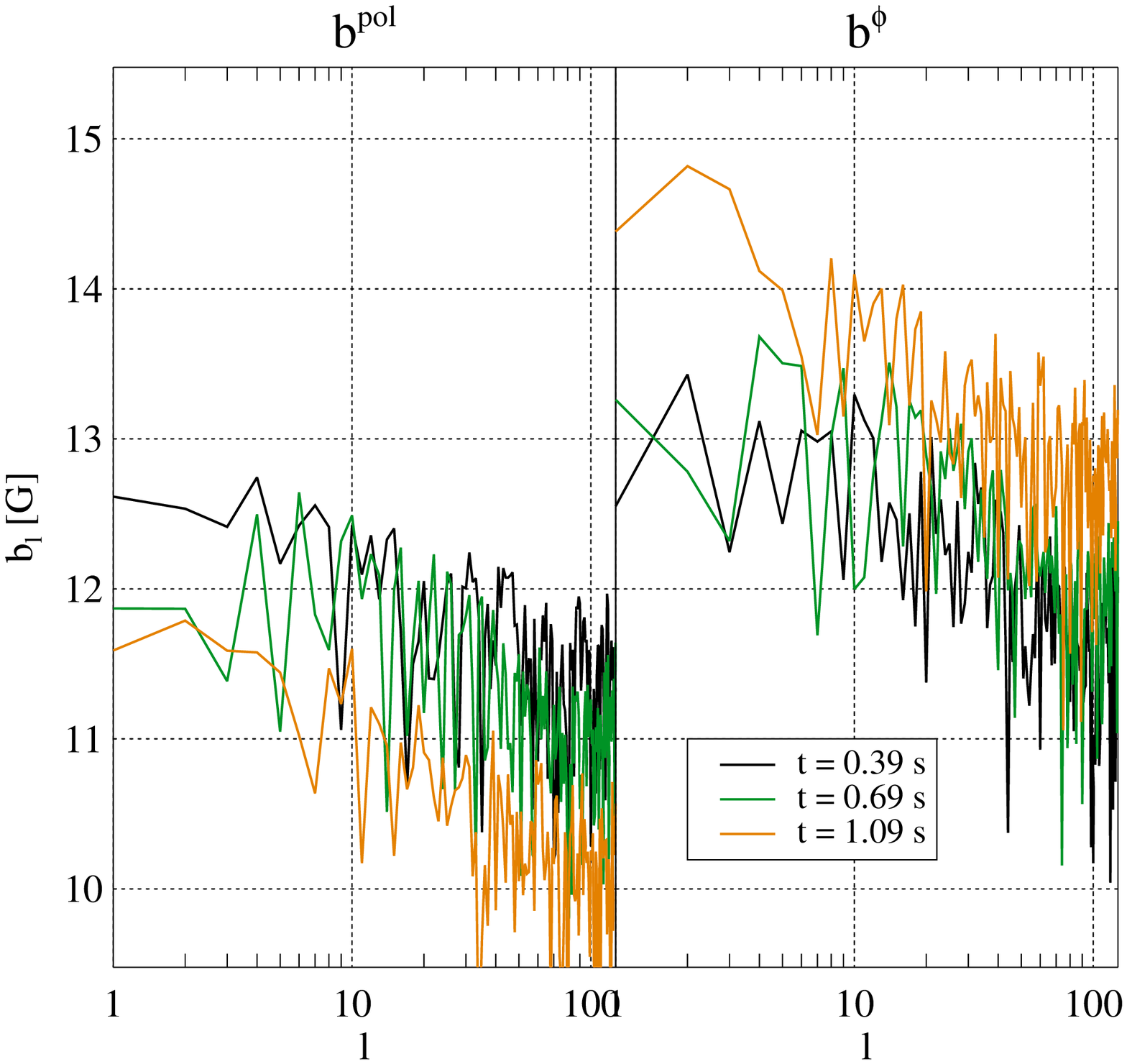}
  \caption{%%
    Top row: maps of the current density (colours in the left parts of
    the figures) and toroidal field component (right parts) overlayed
    by field lines of models (left to right) \modelname{35OC-RO},
    \modelname{35OC-Rs}, and \modelname{35OB-RO}.  For each model, we
    only show the surface layer of the PNS at three different times as
    printed in the top left part of the figures.  
    Due to the contraction of the PNS, the outermost,
    intermediate, and innermost PNS surfaces correspond to the sequence
    from the earliest to the latest times.
    Bottom row: angular spectra of the absolute value of the poloidal
    and the toroidal field components of the same models at the three
    times shown in the to row.
  }
  \label{Fig:35OC-RO-PNSsurf}
\end{figure}

We illustrate the evolution of the surface field of the three models
in \figref{Fig:35OC-RO-PNSsurf}.  At three selected times between an
early stage after bounce and one late in the evolution (close to BH
collapse for \modl{35OB-RO}), we identified the PNS surface with the
angle-dependent radius of $\nu_e$-sphere, $R_{\nu} (\theta)$ and
focussed on a region centred at this radius, $r(\theta) \in
[R_{\mathrm{min}}; R_{\mathrm{max}}] = [R_{\nu}(\theta) - \Delta R; R_{\nu}(\theta) +
\Delta R]$ with $\Delta R = 4 \, \km$.  The top row of the figure
shows the current density, $j = \left( \vec \nabla \times \vec b
\right)^{\phi}$, and the toroidal field strength together with
(poloidal) field lines in this region.  Since the radii of the PNS
surface decreases as time progresses, the three areas in each panel
possess an ordering in time with the outermost and innermost being the
first and last of the three times, respectively.  In addition, we
computed the angular spectra of the field at the same times (bottom
row of panels) by expanding its volume-averaged poloidal and toroidal
components in terms of spherical harmonics $Y_{lm}$ ($m = 0$ because
of axisymmetry):
\begin{eqnarray}
  \label{Gl:SpherHarm}
  b_l^{\mathrm{pol},\phi}&  = & \int \mathrm{d}  \theta \,
  Y_{lm} (\theta)
  \tilde{b}^{\mathrm{pol},\phi} (\theta),
  \\ 
  \label{Gl:btilde}
  \tilde{b}^{\mathrm{pol},\phi} (\theta) 
  & = & \int \mathrm{d} V  \,
  b^{\mathrm{pol},\phi} (r,\theta) 
  \left( \int \mathrm{d} V \right)^{-1},
\end{eqnarray}
where the volume integration extends over the surface region defined above.

We first note that no crust has formed in any of the models.  Hence,
the dynamics of the surrounding region in which the PNS is immersed,
characterized by radial as well as angular flows, leaves an imprint on
the structure and evolution of the surface field.  Consequently, we
can find important variations of the strength and orientation of the
field on intermediate length scales at all times for \modl{35OC-RO}
(left panels).  At $t = 0.24 \, \sek$, regions where the poloidal
field is mostly radial are situated close to ones where the
$\theta$-component is strongest, and the toroidal field shows
reversals of polarity, though on average negative/positive in the
northern/southern hemisphere.  These patterns are reflected in rather
strong current localized in domains of an angular extent around
$\lesssim 20$ degrees.  These general observations hold also at later
times.  In addition, we find an enhancement of the field strength at
the north and south poles, where the PNS radius is lowest.  There, the
poloidal field is mostly radial.  Variations on very small length
scales lead to intense currents.  We note that the gas ending up in
the bipolar explosion is accelerated in the small volume immediately
outside the polar PNS surface.  

The angular spectra of the poloidal field are rather similar for all
three times shown here, except for a growth of the mean field
strength.  Modes up to $l \lesssim 10$ possess a similar spectral
power.  Going to higher orders, the spectra decay according to a power
law with an index slightly exceeding 1.  The spectra of the toroidal
field are somewhat steeper towards low order modes with a power-law
component beginning already at $l < 10$.  In agreement with the
partition of the energy between poloidal and toroidal field, the
toroidal spectrum is about an order of magnitude stronger than the
poloidal one.

\Modl{35OC-Rs} presents a similar case, albeit at a much higher axis
ratio of the PNS and with a more equal distribution of energies of the
poloidal and toroidal components.  Compared to the above model, the
enhancement of the radial field and the concentration of the currents
at the polar caps are more pronounced.  As before,  the poloidal
component shows a spectral plateau at $l \lesssim 10$, while the
spectrum of the toroidal field has a power-law shape across the entire
range of modes.

The evolution of \modl{35OB-RO} differs from this pattern mostly due
to the large unmagnetized shell that is accreted onto the PNS after
the iron core.  At an early stage of the model ($t = 0.39 \, \sek$), the
qualitative results are similar to the other two models: variations on
large and intermediate scales dominate the structure of both field
components and the surface layer is filled by a highly variable
pattern of currents.  However, at later times, gas without a magnetic
field is accreted, burying the field already present in the PNS.
Therefore, the PNS surface at $t = 0.69 \, \sek$ is, except for the
polar caps, almost devoid of magnetic field.  At intermediate to low
latitudes, poloidal field can be found only at its base.  It does not
reach the matter surrounding the PNS.  The toroidal field is slightly
less confined to the interior, but also rather weak outside the PNS.
During the later phases of accretion and contraction of the PNS, this
picture is modified slightly.  At $t = 1.09 \, \sek$, the structure of
the field is visible to a larger extent at the bottom of the surface
layer.  The poloidal field lines are now entirely closed within the
PNS.  Even the polar region where at $t = 0.69 \, \sek$ still some
radial field lines were present is no longer threaded by field lines
connected to the surrounding matter.  Still, the toroidal field is
slightly more extended, but also drops quickly with radius.  Any
speculation as to whether this mostly buried magnetic field might
reemerge later is idle because of the quick collapse to a BH.  For
this model, the plateaus in the spectra of the poloidal component tend
to be even more extended than for the two models discussed above.  The
toroidal component, which is by far stronger than, also shows notable
power at high-order modes.

\section{Conclusions}
\label{Sek:Con}

We investigated the structure and evolution of the magnetic field on
the surface of rapidly rotating PNSs, which are the results of the
axisymmetric simulations coupling magnetohydrodynamics with a spectral
two-moment neutrino transport solver of
\cite{Obergaulinger_Aloy__2017__mnras__Protomagnetarandblackholeformationinhigh-massstars}.
We selected three representative models based on two different cores,
one developing a magnetically supported explosion and leaving behind a
PNS (\modelname{35OC-RO}), one exploding very energetically driven by
the magnetic field and with an extremely oblate PNS at the centre
(\modelname{35OC-Rs}), and a third one that explodes with little
magnetic influence and produces a BH at the end of the simulation
(\modelname{35OB-RO}).

For each of the models, we studied the morphology of the poloidal and
toroidal magnetic field components on the PNS surface, which we
approximately identified with the $\nu_e$-sphere, at different times
and computed the angular spectra of the two components by expanding
their distribution in terms of spherical harmonics.  We typically find
a rather complex field geometry with field lines tangled on small to
intermediate length scales.  An enhancement of the field strength at
the polar caps of the PNSs is particularly prominent for
\modls{35OC-RO} and \modelname{35OC-Rs}.  For both models, the surface
fields are quite strong, reaching or exceeding $10^{14} \, \Gauss$ on
average, which would correspond to magnetars when placed in the
parameter space of old neutron stars.  The angular spectra of the
field can be described in terms of a combination of a plateau at low
mode orders, $l$, and a power-law decay towards higher $l$.  The
former component is most clearly visible for the poloidal component,
for which it extends to $l \gtrsim 10$, whereas the toroidal component
shows only slight hints of its presence at lowest $l$.  This behaviour
means that the field cannot be described well in terms of a single
low-order mode such as, \eg a dipole.  \Modl{35OB-RO} possesses a
similar field early on, but later undergoes a transition to a very
weakly magnetized surface as the magnetic field is buried underneath
unmagnetized gas accreted at late times.  The spectra of the field
shows the same two components, but on a much weaker strength compared
to the other two models strength.

We attribute this field structure to two main agents governing the
field evolution: the internal dynamics of the PNS and the accretion of
matter with a varying degree of magnetization.  The PNSs develop
convective activity quickly after their formation.  While the surface
usually is not part of the convective layer, it is nevertheless
affected by overshoot from below.  Convective eddies typically have
sizes up to intermediate angles of few tens of degrees, which are
reflected in the strength of the spectra at these angles.  The field
embedded in the gas falling onto the PNS surface is modulated by the
non-radial flows in the post-shock layer on angular scales similar to
those of the motions inside the PNS.  At late times, after the onset
of the explosions, these accretion flows take the shape of relatively
narrow streams with an angular width that also corresponds to
intermediate orders $l \sim 10$ impinging on the PNS at stochastically
changing locations.  Hence, the effect of both processes cannot be
clearly disentangled.  In the absence of a magnetic field in the
accreted matter, the PNS surface rather quickly turns virtually
non-magnetic, despite the ongoing activity of magneto-convection
below.  This fact suggests that accretion rather than the PNS
convection is the process mainly responsible for generating and
maintaining the surface field.

The main limitations of the current study, both caused by the
restrictions of computational time, are the assumption of axisymmetry
and the still fairly limited simulation time.  The latter restriction
prevents us from studying, \eg the possible reemergence of buried
magnetic fields in cores obtaining a similar structure to
\modl{35OB-RO}.  To remedy both drawbacks at the same time is
currently not feasible.  Shorter three-dimensional runs, on the other
hand, can be done at reasonable costs, albeit not for a large number
of models.  Running the simulations until the crust formation would be
desirable since the field evolution on the surface is slowed down
considerably after that moment.  So far, such a long simulation time
of about 70\,s has been achieved by
\cite{Suwa__2014__pasj__Fromsupernovaetoneutronstars} in spherical
symmetry.  To reach comparable times in axisymmetry, we would have to
drastically reduce the computational costs by replacing the full
neutrino transport by a much simpler treatment at a suitably chosen
moment.  Otherwise, computational costs range in several millions of
core-hours per model.  This problem is exacerbated by the fact that
the main factor driving up the costs is the not the grid size, but the
large number of time steps, against which parallelization does not
offer a straightforward solution.

\section*{Acknowledgements}
\label{Sek:Ackno}

We acknowledge support from the European Research Council (grant
CAMAP-259276) and from the Spanish Ministry of Economy and Finance and
the Valencian Community grants under grants AYA2015-66899-C2-1-P and
PROMETEOII/2014-069, resp.  We thank Oliver Just and Thomas Janka for
valuable help, in particular regarding the aspects of microphysics and
neutrinos. 
The computations were
performed under grants AECT-2016-1-0008, AECT-2016-2-0012,
AECT-2016-3-0005, AECT-2017-1-0013, and AECT-2017-2-0006 of the
Spanish Supercomputing Network on hosts \textit{Pirineus} of the
Consorci de Serveis Universitaris de Catalunya (CSUC),
\textit{Picasso} of the Universidad de M{\'a}laga,
\textit{FinisTerrae2} of the Centro de Supercomputaci{\'o}n de
Galicia, Santiago de Compostela, and \textit{MareNostrum} of the
Barcelona Supercomputing Centre, respectively, and on the clusters
\textit{Tirant} and \textit{Lluisvives} of the Servei d'Inform\`atica
of the University of Valencia.

\bibliographystyle{mn2e}

\end{document}